\documentclass[preprint,12pt,nofootinbib,groupedaddress,superscriptaddress]{revtex4-2}

\usepackage{graphicx}
\usepackage{dcolumn} 
\usepackage{bm}
\usepackage{amssymb}
\usepackage{amsmath}
\usepackage{epsfig}    
\usepackage{color}
\usepackage{slashed}
\usepackage{hyperref}

\begin{document}

\title{Multi-$Z'$ signatures of spontaneously broken local $U(1)'$ symmetry}
\preprint{OU-HET-1235}

\author{Takaaki Nomura}
\email{nomura@scu.edu.cn}
\affiliation{College of Physics, Sichuan University, Chengdu 610065, China}

\author{Kei Yagyu}
\email{yagyu@het.phys.sci.osaka-u.ac.jp}
\affiliation{Department of Physics, Osaka University, Toyonaka, Osaka 560-0043, Japan}

\begin{abstract}
\noindent
We discuss multi-$Z'$ signatures coming from decays of Higgs bosons in models with a spontaneously broken $U(1)'$ symmetry, which can be observed as ``lepton jets" or multi-lepton final states depending on the mass range of new bosons.  
We consider anomaly-free $U(1)'$ models without introducing new fermions except for right-handed neutrinos, in which the Higgs sector is composed of an isospin doublet and a singlet fields with zero and non-zero $U(1)'$ charges, respectively. 
The multi-$Z'$ signatures can then be obtained via the decays of the discovered (extra) Higgs boson $h$ ($\phi$), i.e., $h\to Z'Z'$, $\phi \to Z'Z'$ and/or $h \to \phi\phi \to 4Z'$ as far as kinematically allowed.
We give the upper limit on the branching ratios of $h$ into $Z'Z'$ and $4Z'$ from the current experimental data in each model. 
We also show the deviation in the $hhh$ coupling from the standard model prediction at one-loop level, and find that its amount is typically smaller than 1\%.  
\end{abstract}

\maketitle
\newpage


\section{Introduction}

The electroweak (EW) gauge symmetry and its spontaneous breakdown have been confirmed by various experiments to date.
In particular, the discovery of the Higgs boson at LHC has revealed the realization of the spontaneous symmetry breaking via the Higgs mechanism. 
Although there is no contradiction between 
the structure of $SU(2)_L \times U(1)_Y \to U(1)_{\rm em}$ implemented in the Standard Model (SM) and current experimental data, one can regard such a pattern of the symmetry breaking as a part of larger structures. 

Models with a new $U(1)$ gauge symmetry, denoting $U(1)'$, have often been discussed as one of the simplest and well-motivated examples of such an extended EW symmetry.  In fact, grand unified theories such as $SO(10)$ and $E_6$ models predict extra $U(1)$ gauge symmetries at the EW scale, see e.g., Ref.~\cite{Hewett:1988xc} and references therein. 
In addition, gauging the $U(1)_{B - L}$ symmetry~\cite{Pati:1973uk,Davidson:1978pm,Marshak:1979fm} can be regarded as a natural extension of the EW symmetry because the global $U(1)_{B - L}$ accidentally appears in the SM. 
In models with the gauged $U(1)_{B - L}$ symmetry, three right-handed neutrinos are inevitably introduced to maintain the gauge anomaly cancellation, and they make left-handed neutrinos massive via the seesaw mechanism~\cite{Minkowski:1977sc,Yanagida:1979as,Gell-Mann:1979vob,Mohapatra:1979ia}. 
More generally, one can consider the $U(1)_X$ gauge symmetry~\cite{Appelquist:2002mw,Das:2017flq} whose 
charges of SM fields are expressed by two free parameters, and the well-known examples, e.g., $U(1)_{B - L}$, $U(1)_D$ (dark photon)~\cite{Holdom:1985ag} and $U(1)_R$ (right-handed)~\cite{Ko:2012hd,Nomura:2017tih} correspond to its special cases. Furthermore, if we allow a flavor dependence of the charge assignments of $U(1)'$, three possibilities, i.e., $U(1)_{L_i-L_j}$ ($L_i$ and $L_j$ being the lepton number of $i$-th and $j$-th flavor, respectively)~\cite{Foot:1990mn,He:1990pn}, appear under the anomaly free condition without introducing any exotic fermions\footnote{One can also consider $U(1)_{Q_i-Q_j}$ symmetries as the anomaly free choice, where $Q_i$ and $Q_j$ are the quark number of $i$-th and $j$-th flavor, respectively. However, these models predict a diagonal form of the quark Yukawa matrices for the case with one Higgs doublet without charged under $U(1)_{Q_i-Q_j}$, and they cannot explain the CKM matrix. }. In particular, models with $U(1)_{L_\mu-L_\tau}$ have often been discussed as a solution of the muon $g-2$ anomaly~\cite{Baek:2001kca,Ma:2001md,Baek:2015mna}. 
As the other possibilities, one can also consider linear combinations of $U(1)_{B - L}$ and $U(1)_{L_i-L_j}$ symmetries, e.g., $U(1)_{B - 3L_i}$ symmetries. 
Therefore, it is important to comprehensively investigate scenarios with $U(1)'$ to determine new physics beyond the SM.  

A common feature of models with $U(1)'$ is the appearance of an extra neutral gauge boson $Z'$. 
We discuss a massive $Z^\prime$ case because 
a massless $Z'$ has highly been constrained by tests of violation of the weak equivalence theorem~\cite{Wagner:2012ui}, which give the bound on the new gauge coupling constant, e.g., in the $U(1)_{B-L}$ case, to be smaller than $10^{-24}$~\cite{Heeck:2014zfa}. 
Since we have already confirmed the spontaneous breaking of the EW symmetry via the Higgs mechanism, it is quite natural to consider that the $U(1)'$ symmetry is also spontaneously broken by the Higgs mechanism.~\footnote{Alternatively, one can introduce the $Z'$ mass via the Stueckelberg mechanism~\cite{Stueckelberg:1938hvi,Ruegg:2003ps} without introducing Higgs fields while keeping the gauge invariance. 
However, this is nothing but a special case of the Higgs mechanism where a new physical Higgs boson is decoupled from the theory and the so-called Stueckelberg scalar field can be regarded as a Nambu-Goldstone boson associated with the spontaneous $U(1)'$ breaking. } 
The simplest way to do that is to introduce a complex scalar field which is singlet under the SM gauge symmetry.  
In this class of models, an additional physical neutral Higgs boson $\phi$ also appears in addition to $Z'$, and their masses are naturally expected to be given in a similar order because they come from the Vacuum Expectation Value (VEV) of the singlet scalar. 
Therefore, the phenomenological impacts of two new bosons should be taken into account at the same time.   
In particular, if the masses of $Z'$ and $\phi$ are smaller than the discovered Higgs boson ($h$), 
the latter can decay into a $Z'$ pair and/or $\phi$ pair. 
The $h \to Z'Z'$ mode has been discussed in the dark photon model, i.e., the $U(1)_D$ model, 
in Refs.~\cite{Gopalakrishna:2008dv,Davoudiasl:2013aya,Chang:2013lfa,Curtin:2013fra,Falkowski:2014ffa,Curtin:2014cca}. 
In addition, the decay channel of $h \to \phi\phi \to 4Z' \to 8$ fermions has been discussed in the $U(1)_D$ model~\cite{Chang:2013lfa,Izaguirre:2018atq}. 
Charged leptons produced via the decay of $Z'$ can be observed as ``lepton jets"~\cite{Arkani-Hamed:2008kxc,Falkowski:2010gv} if there is a large mass difference between the discovered Higgs boson and the masses of $Z'/\phi$, especially around GeV or less due to 
highly boosted and collimated charged leptons. 
%
As for the other $U(1)'$ models, 
Ref.~\cite{Nomura:2020vnk} and Ref.~\cite{A:2024shl} have discussed the $Z^\prime$ production via the decay of $\phi$ in the  $U(1)_{L_\mu - L_\tau}$ model and that of $h$ in the $U(1)_X$ model, respectively. 

In this paper, we discuss the phenomenology of models with various $U(1)'$ extensions of the SM with a complex scalar field. 
We focus on the impact of the collider phenomenology in the presence of $Z'$ and $\phi$ having a similar size of the masses, in which we classify three regions of the masses, i.e.,  (i) small mass ${\cal O}(10)$ MeV, (ii) middle mass ${\cal O}(10)$ GeV and (iii) large mass ${\cal O}(100)$ GeV.
We show the branching ratios of the new decay channels of $h$ for the cases (i) and (ii) 
as a function of the new gauge coupling ($g_X^{}$) and the mixing angle of $h$ and $\phi$ ($\sin\alpha$)
in each $U(1)'$ model under the constraints from flavor and collider experiments. 
We then clarify the upper limit on these branching ratios in each model. 
For the case (iii), we show the branching ratio of the new Higgs boson $\phi$ into a $Z'$ pair as a function of $g_X^{}$ and $\sin\alpha$.
In addition to these branching ratios, we also analyze the deviation of the Higgs trilinear coupling $hhh$ from the SM prediction at one-loop level, 
and find that the magnitude of the deviation is typically less than $1\%$ level.


This paper is organized as follows. 
In Sec.~\ref{sec:models}, we define our model with the $U(1)'$ gauge symmetry, and give the Lagrangian and relevant interaction terms. 
In Sec.~\ref{subsec:decay},  we discuss the decays of the $Z'$ boson and the Higgs bosons. 
Sec.~\ref{sec:hhh} is devoted to calculating the one-loop corrections to the $hhh$ coupling. 
In Sec.~\ref{sec:pheno}, we discuss the collider phenomenology of our scenario focusing on the multi-$Z'$ signature coming from the decay of the Higgs bosons. 
Summary and discussions are given in Sec.~\ref{sec:summary}. 
In Appendix~\ref{sec:ren_zp}, we present details of the renormalization which is necessary to calculate the one-loop correction to the $hhh$ coupling.  

\section{Models\label{sec:models}}

\begin{table}[t]
  \begin{center}
    \begin{tabular}{|c|c|c|c|c|c|c|c|c|c|c|c|c|}\hline
      &
      $~~q_L^{i}~~$ & 
      $~~u_R^{i}~~$ & 
      $~~d_R^{i}~~$ & 
      $~~l_L^e~~$ & 
      $~~l_L^\mu~~$ & 
      $~~l_L^\tau~~$ & 
      $~~e_R^{}~~$ & $~~\mu_R^{}~~$ & $~~\tau_R^{}~~$ &
      $~~\nu_{R}^e~~$ & $~~\nu_{R}^\mu~~$ & $~~\nu_{R}^\tau~~$  \\ \hline
      $~~SU(2)_L~~$ & $\bf{2}$ & $\bf{1}$ & $\bf{1}$ & 
      $\bf{2}$ & $\bf{2}$ & $\bf{2}$ & $\bf{1}$ &$\bf{1}$ & $\bf{1}$ & $\bf{1}$ & $\bf{1}$ & $\bf{1}$  \\ \hline
      $~~U(1)_Y~~$ & $\frac{1}{6}$ & $\frac{2}{3}$ & $-\frac{1}{3}$ & 
      				$-\frac{1}{2}$ & $-\frac{1}{2}$ & $-\frac{1}{2}$ & $-1$ &$-1$ & $-1$ & $0$ & $0$ & $0$ \\ \hline
      $~~U(1)'~~$ & $X_{q_L}$ & $X_{u_R}$ & $X_{d_R}$ & $X_{l_L^e}$ & $X_{l_L^\mu}$ & $X_{l_L^\tau}$ & $X_{e_R}$ &$X_{\mu_R}$ & $X_{\tau_R}$ & $X_{\nu_{R}^e}$
        & $X_{\nu_{R}^\mu}$ & $X_{\nu_{R}^\tau}$ \\ \hline
    \end{tabular}
  \end{center}
  \caption{Charge assignments for fermions in a gauged $U(1)'$ model, where charges for quark fields ($q_L^{i}$, $u_R^i$ and $d_R^i$) are taken to be flavor universal. }
  \label{tab:charge}
%
  \begin{center}
    \begin{tabular}{|c|c|c|c|c|c|c|c|c|c|c|c|c|}\hline
 & $X_{q_L}$ & $X_{u_R}$ & $X_{d_R}$ & $X_{l_L^e}$ & $X_{l_L^\mu}$ & $X_{l_L^\tau}$ & $X_{e_R}$ &$X_{\mu_R}$ & $X_{\tau_R}$ & $X_{\nu_{R}^e}$ & $X_{\nu_{R}^\mu}$ & $X_{\nu_{R}^\tau}$ \\ \hline
 $~~U(1)_{B-L}~~$ & $\frac13$ & $\frac13$ & $\frac13$ & $-1$ & $-1$ & $-1$ & $-1$ & $-1$ & $-1$ & $-1$ & $-1$ & $-1$ \\ \hline
 $~~U(1)_{L_e-L_\mu}~~$ & $0$ & $0$ & $0$ & $1$ & $-1$ & $0$ & $1$ & $-1$ & $0$ & $1$ & $-1$ & $0$  \\ \hline
 $~~U(1)_{L_e-L_\tau}~~$ & $0$ & $0$ & $0$ & $1$ & $0$ & $-1$ & $1$ & $0$ & $-1$ & $1$ & $0$ & $-1$ \\ \hline
 $~~U(1)_{L_\mu-L_\tau}~~$ & $0$ & $0$ & $0$ & $0$ & $1$ & $-1$ & $0$ & $1$ & $-1$  & $0$ & $1$ & $-1$ \\ \hline
 $~~U(1)_{D}~~$ & $0$ & $0$ & $0$ & $0$ & $0$ & $0$ & $0$ & $0$ & $0$ & $0$ & $0$ & $0$ \\ \hline
    \end{tabular}
  \end{center}
  \caption{Charges $X_f$ for a fermion $f$ in different $U(1)'$ models. }
  \label{tab:models}
\end{table}

We consider models with an extra $U(1)'$ gauge symmetry in the minimal extension of field contents where we introduce three right-handed neutrinos $\nu_R^i (i=1-3)$ and a complex singlet scalar field $\Phi$.
For simplicity, we take the SM Higgs doublet $H$ to be neutral under $U(1)'$.
Typical candidates of such $U(1)'$ are then $U(1)_{B-L}$, $U(1)_{L_i - L_j}$ and the dark $U(1)_D$ symmetry under the anomaly cancellation condition.
We summarize the $U(1)'$ charges for the SM fermions and right-handed neutrinos in Table~\ref{tab:charge} and Table~\ref{tab:models}.

The Lagrangian is generally written by
\begin{align}
\mathcal{L} =& \mathcal{L}_{\mathrm{SM}}- \frac{1}{4} X_{\mu \nu} X^{\mu \nu}  - \frac{\epsilon}{2} B_{\mu \nu} X^{\mu \nu}  
+ g_X X_\mu J^\mu_{X}  +|D_\mu \Phi|^2 - V(H,\Phi) + \mathcal{L}_{\nu_R},
 \label{eq:Lagrangian} 
\end{align}
where $\mathcal{L}_{\mathrm{SM}}$ denotes the SM Lagrangian without the Higgs potential, 
$X_{\mu\nu}$ ($B_{\mu\nu}$) is 
the field strength tensor for the $U(1)'$ (hypercharge) gauge field $X_\mu$  ($B_\mu$), $J_\mu^X$ is the current associated with $X^\mu$, 
$D_\mu \Phi \equiv (\partial_\mu + ig_X^{}X_\Phi X_\mu)\Phi$ is the covariant derivative for $\Phi$, $V$ is the Higgs potential, and $\mathcal{L}_{\nu_R}$ represents terms with right-handed neutrinos. 
In Eq.~(\ref{eq:Lagrangian}), we introduced 
$\epsilon$ and $g_X^{}$ denoting the kinetic mixing parameter and the gauge coupling of $U(1)'$, respectively.

Let us comment on the last term $\mathcal{L}_{\nu_R}$ which includes the kinetic term for $\nu_R^i$ and the Yukawa interaction among $H$, $l_L^i$ and $\nu_R^i$. 
Depending on the choice of the $U(1)'$ symmetry and its charge assignments, $\mathcal{L}_{\nu_R}$ can also contain Majorana masses of $\nu_R^{i}$ and/or the Yukawa interaction with $\Phi$ which generates the Majorana mass after the spontaneous breaking of $U(1)'$. Such a Majorana mass turns out to be masses of active neutrinos via the type-I seesaw mechanism. 
In the following discussion, we mainly focus on the phenomenology of the Higgs and $Z'$ sector, and details of the neutrino sector do not affect on the analyses given below. 
For a concrete analysis, we fix $X_{\Phi} = 1$ for all the $U(1)'$ models considered in this paper. 

\subsection{Higgs potential and Higgs boson masses} \label{subsec:scalar-sector}

Here, we formulate eigenvalues for scalar bosons and corresponding eigenstates after the spontaneous gauge symmetry breaking.

The Higgs potential $V$ is generally written by 
\begin{align}
V(H,\Phi) =   -\mu_H^2 H^\dagger H - \mu_\Phi^2 \Phi^* \Phi + \frac{\lambda_H}{2} (H^\dagger H)^2 + \frac{\lambda_\Phi}{2} (\Phi^* \Phi)^2 + \lambda_{H \Phi} (H^\dagger H)(\Phi^* \Phi), \label{eq:scalar-potential}
\end{align} 
where $\mu_H^{}$ and $\mu_\Phi^{}$ are the mass parameters, and $\lambda_H,~\lambda_\Phi$ and $\lambda_{H\Phi}$ are the quartic couplings, 
respectively. 
We can parameterize the $H$ and $\Phi$ fields as 
\begin{align}
H = 
\begin{pmatrix}
G^+ \\
\frac{1}{\sqrt{2}} (v + \tilde{h} + i G) 
\end{pmatrix}, \quad
\Phi &= \frac{1}{\sqrt{2}} (v_\Phi + \tilde{\phi} + i G_\Phi),
\label{eq:scalar-fields}
\end{align}
where $\tilde{h}$ and $\tilde{\phi}$ denote physical CP-even scalar components while $G^\pm$, $G$ and $G_\Phi$ are the Nambu-Goldstone bosons that are absorbed by the weak gauge bosons $Z,~W$ and extra $Z'$ boson, respectively. 
The mass squared matrix of the CP-even scalar bosons is given by
\begin{align}
M^2_{\mathrm{even}} = 
\begin{pmatrix}
\lambda_H v^2 & \lambda_{H\Phi} v v_\Phi \\
\lambda_{H\Phi} v v_\Phi & \lambda_\Phi v_\Phi^2
\end{pmatrix},
\end{align}
where we adopted the stationary conditions,
\begin{align}
\frac{\partial V}{\partial \tilde h}\Bigg|_0 = \frac{\partial V}{\partial \tilde \phi}\Bigg|_0 =0, 
\end{align}
with $|_0$ denoting all the component fields to be zero. 
Then, we can diagonalize the mass matrix by introducing an orthogonal matrix $O_{\mathrm{even}}$,
\begin{align}
 O^T_{\mathrm{even}} M^2_{\mathrm{even}} O_{\mathrm{even}} = \mathrm{diag}(m_h^2, m_\phi^2),
\end{align}
where 
\begin{align}
O_{\mathrm{even}} = 
\begin{pmatrix}
\cos\alpha & -\sin\alpha \\
\sin\alpha & \cos\alpha
\end{pmatrix}.
\end{align}
The mass eigenvalues are also given by
\begin{subequations}
\begin{align}
m_h^2 &= \lambda_H v^2 \cos^2\alpha + \lambda_\Phi  v_\Phi ^2 \sin^2\alpha 
	+ 2 \lambda_{H\Phi} v v_\Phi \sin\alpha \cos\alpha, \\
m_\phi^2 &= \lambda_\Phi v_\Phi^2 \cos^2\alpha + \lambda_H v^2 \sin^2\alpha 
	- 2 \lambda_{H\Phi} v v_\Phi \sin\alpha \cos\alpha. 
\end{align}
\label{eq:scalar-masses}
\end{subequations}
The scalar mixing angle $\alpha$ is obtained as 
\begin{align}
\tan 2\alpha = \frac{2 \lambda_{H \Phi} v v_\Phi}{\lambda_H v^2 - \lambda_\Phi v_\Phi^2}. \label{eq:scalar-mixing}
\end{align}
The corresponding mass eigenstates are written by
\begin{align}
\begin{pmatrix}
h \\
\phi
\end{pmatrix}
=
O^T_{\mathrm{even}}
\begin{pmatrix}
\tilde{h} \\
\tilde{\phi}
\end{pmatrix}. 
\label{eq:scalar-eigenstates}
\end{align}
where 
we identify $h$ with the discovered Higgs boson with a mass of 125 GeV. 
From Eqs.~\eqref{eq:scalar-masses} and \eqref{eq:scalar-mixing}, the quartic couplings can be 
expressed in terms of the masses and the mixing as
\begin{subequations}
\begin{align}
\lambda_H &= \frac{1}{2 v^2}\big[ m_h^2 + m_\phi^2 + (m_h^2 - m_\phi^2) \cos 2\alpha \big], \\
\lambda_\Phi &= \frac{1}{2 v_\Phi^2}\big[ m_h^2 + m_\phi^2 - (m_h^2 - m_\phi^2) \cos 2\alpha \big], \\
\lambda_{H\Phi} &= \frac{1}{2 v v_\Phi} (m_h^2 - m_\phi^2) \sin 2\alpha.
\end{align} 
\label{eq:quartic_coup}
\end{subequations}
Notice that $\lambda_H$ and $\lambda_\Phi$ are always positive due to positive mass squared of $h$ and $\phi$ 
while $\lambda_{H\Phi}$ can be negative.

The trilinear couplings for the scalar bosons are then expressed in their mass bases as follows
\begin{align}
\label{eq:intV}
\mathcal{L} \supset & \left(\frac{m_Z^2}{v}  Z_\mu Z^\mu + \frac{2 m_W^2}{v} W^+_\mu W^{-\mu}-\sum_{ f_{\rm SM}} \frac{m_{f_{\rm SM}}}{v} \overline{ f_{\rm SM}} f_{\rm SM}  \right) ( h\cos \alpha - \phi\sin \alpha) \nonumber \\
&  + \frac{m_{Z'}^2}{v_\Phi}  Z'_\mu Z'^\mu (\phi\cos \alpha  +h \sin \alpha ) +\sum_{\rm scalars}\lambda_{\varphi_i\varphi_j\varphi_k}\varphi_i\varphi_j\varphi_k,  
\end{align}
where $m_{f_{\rm SM}}$  represents the mass of the SM fermions $f_{\rm SM}$ ($f_{\rm SM} \neq \nu_R$). The last term shows the scalar trilinear terms, and the relevant couplings are given as follows: 
\begin{align}
\lambda_{hhh} &= -\frac{m_h^2(v_\Phi \cos^3 \alpha  + v\sin^3 \alpha )}{2v v_\Phi}, \label{eq:hhh} \\
\lambda_{\phi hh} &= \frac{\sin 2\alpha (2 m_h^2 + m_\phi^2) (v_\Phi \cos \alpha  - v\sin \alpha )}{4 v v_\Phi}, \label{eq:phihh}\\
\lambda_{\phi \phi h} &=- \frac{\sin 2\alpha (m_h^2 + 2 m_\phi^2) (v \cos \alpha  +v_\Phi \sin \alpha )}{4 v v_\Phi}, \label{eq:phiphih}\\
\lambda_{G_\Phi G_\Phi h} &= -\frac{m_h^2\sin\alpha}{2v_\Phi}.
\end{align}

\subsection{$Z'$ mass and interactions} \label{subsec:gauge-sector}

Because of the existence of the kinetic mixing term in Eq.~\eqref{eq:Lagrangian}, we first need to diagonalize the
kinetic terms of $X_\mu$ and $B_\mu$ by the transformation 
\begin{equation}
\left(\begin{array}{c}
X^3_\mu\\
B_\mu\\
\end{array}\right)=\left(\begin{array}{cc}
r & 0 \\
-\epsilon r & 1 \\
\end{array}\right)\left(\begin{array}{c}
\tilde{Z}^\prime_\mu\\
\tilde{B}_\mu\\
\end{array}\right),
\label{eq:kinetic}
\end{equation}
where $r = 1/\sqrt{1 - \epsilon^2}$.
After the spontaneous breaking of the EW and the $U(1)'$ gauge symmetries, the mass terms of electrically neutral gauge fields are induced as follows
\begin{align}
\mathcal{L}_{M} =  \frac{1}{2}
(\tilde{Z}_\mu , \tilde{Z}_\mu^\prime)
\begin{pmatrix}
m^2_{Z_{\rm SM}} & \epsilon r \sin \theta_W m^2_{Z_{\rm SM}}\\
\epsilon r \sin \theta_W m^2_{Z_{\rm SM}} &
r^2 \left(g_X^2 v_\Phi^2 +\epsilon^2 m^2_{Z_{\rm SM}} \sin^2 \theta_W  \right)
\end{pmatrix}
\begin{pmatrix}
\tilde{Z}^\mu  \\
\tilde{Z}^{\prime\mu}
\end{pmatrix}, 
\end{align}
%
where $\tilde{Z}_\mu = \cos \theta_W W^3_\mu - \sin \theta_W \tilde{B}_\mu$ 
and 
$ m^2_{Z_{\rm SM}} = v^2g^2/(4\cos^2\theta_W)$
with $\theta_W$ and $g$ being the Weinberg angle and the $SU(2)_L$ gauge coupling, respectively. 
The above mass matrix can be diagonalized by introducing the following orthogonal transformation:  
\begin{align}
\begin{pmatrix}
\tilde{Z}_\mu \\
\tilde{Z}_\mu^\prime 
\end{pmatrix}
=
\begin{pmatrix}
\cos\chi & \sin\chi \\
-\sin\chi & \cos\chi
\end{pmatrix}
\begin{pmatrix}
Z_\mu \\
Z_\mu^\prime 
\end{pmatrix}, 
\end{align}
where $Z_\mu$ can be identified with the observed $Z$ boson. 
The mixing angle $\chi$ is written by 
\begin{align}
\sin 2\chi =  - \frac{2 \epsilon r \sin\theta_W m_{Z_{{\rm SM}}}^2}{m_Z^2 - m_{Z'}^2}.
\label{eq:tan2chi}
\end{align}
Taking a tiny kinetic mixing parameter $\epsilon \ll 1$, deviation of the $Z$ boson mass from the SM value is negligible and
 the $Z'$ boson mass is written by
\begin{align}
m_{Z'} & =  g_X v_\Phi [1 + \mathcal{O}(\epsilon^2)].
\label{eq:gauge-masses}
\end{align}

The gauge current $J_{X}^\mu$ is given by
\begin{align}
J^\mu_{X} = \sum_{f_{}} X_{f_{}} \bar{f}_{} \gamma^\mu f_{}, \label{eq:current} 
\end{align}
where $f_{}$ indicates the fermions in the models.
The relevant interaction Lagrangian for $Z'$ and fermions $f$ is given by
\begin{align}
\mathcal{L}_{Z'f \bar f} =g_X^{}\bar{f} \gamma^\mu (v'_f - a'_f \gamma_5) f Z'_\mu,
\label{eq:LZpff}
\end{align}
where 
\begin{subequations}
\begin{align}
v'_f &= X_{f} r \cos\chi
+ \frac{e}{g_X^{}} Q_f \epsilon r \cos\theta_W \cos\chi \notag\\
&- \frac{g}{g_X^{}\cos\theta_W}\left(\frac{T_{3f}}{2} - Q_f \sin^2\theta_W\right) (\sin\chi + \epsilon r \sin\theta_W \cos\chi), \\
a'_f &= -\frac{g}{g_X^{}\cos\theta_W}\frac{T_{3f}}{2} (\sin\chi + \epsilon r \sin\theta_W \cos\chi),
\end{align}
\label{eq:coup-Zprime-ff}
\end{subequations}
where $T_{3f}$ is the diagonal $SU(2)_L$ generator, $Q_{f}$ is the electric charge, and $X_f$ is the $U(1)'$ charge given in Table~\ref{tab:models}, respectively. 
We note that the dark photon ($X_f =0$) interaction 
is simply proportional to the charge $Q_f$, i.e., 
$\mathcal{L}_{Z' ff} \simeq e Q_f \epsilon r \cos \theta_W \bar f \gamma^\mu f Z'_\mu$ 
for $m_{Z'} \ll m_Z$.



\section{Decays of the Higgs bosons and $Z'$} \label{subsec:decay}

We discuss the decays of the Higgs bosons $\phi$ and $h$ as well as those of $Z'$.

First, the decay rates of $\phi \to Z' Z'$ and $\phi \to hh$ are given by 
\begin{subequations}
\begin{align}
\Gamma(\phi \to Z'Z') &= \frac{m_{Z'}^4 \cos^2\alpha}{8 \pi v_\Phi^2 m_\phi}\beta(x_{Z'}) 
\left[ 2 + \frac{1}{4 x_{Z'}^2} \left( 1 -2x_{Z'}^{} \right)^2 \right], \\\label{eq:phi-decay-zpzp}
\Gamma(\phi \to hh) &= \frac{\lambda_{ \phi h h}^2}{8 \pi m_\phi} \beta(x_h), 
\end{align}
\end{subequations}
%
where $x_i = m_i^2/m_\phi^2$ and  $\beta(x) = \sqrt{1 - 4 x}$. 
The scalar trilinear coupling $\lambda_{\phi hh}$ is given in Eq.~(\ref{eq:phihh}).  
%
The decay rates of $\phi$ 
into a pair of SM particles (except for $h$)
are obtained 
from the corresponding expression of the SM Higgs boson by multiplying the factor of $\sin^2 \alpha$ and 
replacing $m_h \to m_\phi$. 

Second, the decay rates for the $h \to Z' Z'$ and $h \to \phi \phi$ modes are given by
\begin{subequations}
\begin{align}
\Gamma(h \to Z' Z') &= \frac{m_{Z'}^4 \sin^2 \alpha}{8 \pi v_\Phi^2 m_h}  \beta(z_{Z'})
 \left[ 2 + \frac{1}{4 z_{Z'}^2} \left( 1 -
 2z_{Z'}^{} \right)^2 \right], \\
\Gamma(h \to \phi \phi) &= \frac{\lambda_{ \phi \phi h}^2}{8 \pi m_h} \beta(z_\phi), \label{decay:hphiphi}
\end{align} \label{decay:hphiphi2}
\end{subequations}
where $z_i = m_i^2/m_h^2$ and $\lambda_{\phi\phi h }$ is given in Eq.~(\ref{eq:phiphih}).
The other decay modes of $h$ into SM particles are the same as the SM ones except for the overall extra factor of $\cos^2 \alpha$. 
Thus, the branching ratios of $h$ into a pair of SM particles do not change so much from the SM predictions as long as the new decay channels given in Eq.~(\ref{decay:hphiphi2}) do not become large. 

The decay rates of $Z'$ into a fermion pair are calculated as
\begin{align}
\Gamma(Z' \to f \bar{f}) = g_X^2\frac{m_{Z'}}{12 \pi} \beta \left(\frac{m_f^2}{m_{Z'}^2}\right) 
  \left[ {v'_f}^2 + {a'_f}^2 + 2({v'_f}^2 - 3 {a'_f}^2) \frac{m_f^2}{m_{Z'}^2}\right], \label{eq:gamma-Zp-ff}
\end{align}
where $v'_f$ and $a'_f$ are given by Eq.~\eqref{eq:coup-Zprime-ff}.

\begin{figure}[tb]
\begin{center}
\includegraphics[width=70.0mm]{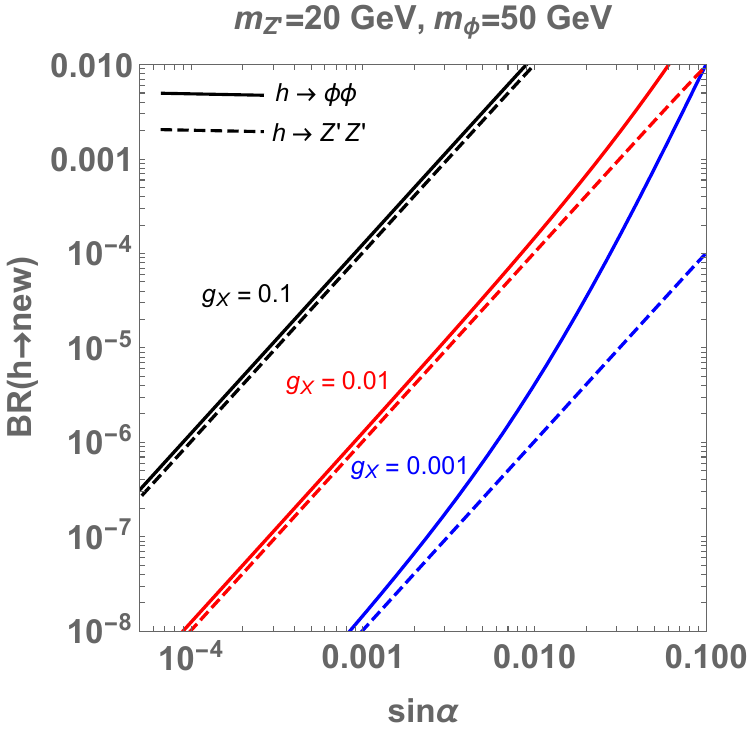} \quad
\includegraphics[width=70.0mm]{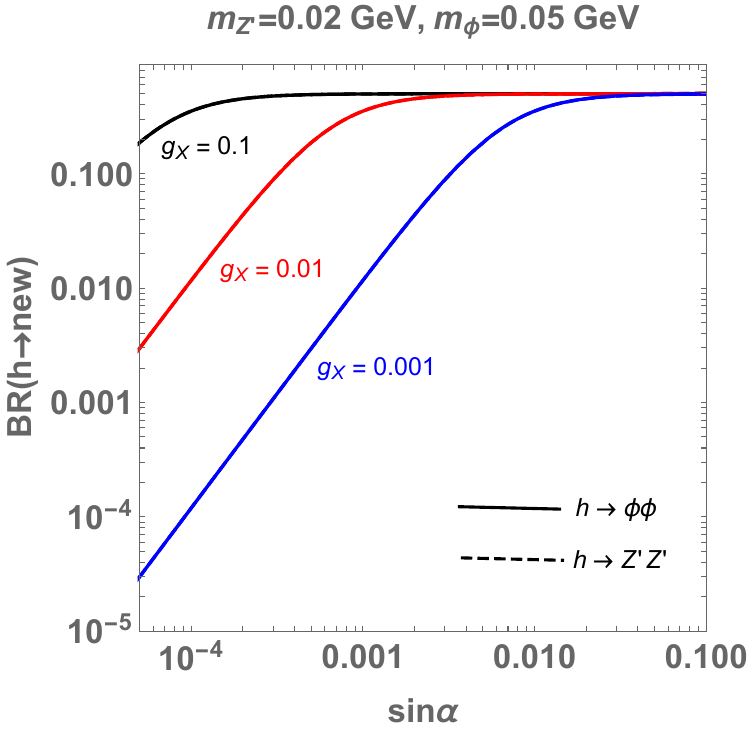}  
\caption{Branching ratios for the Higgs boson decay $h \to Z'Z'$ and 
$h \to \phi \phi$ as a function of $\sin \alpha$ where the masses of $Z'$ and $\phi$ are fixed to be 20 (0.02) GeV and 50 (0.05) GeV in the left (right) plot. We also change $g_X$ as 0.1, 0.01 and 0.001 for each plot as indicated by curves with black, red and blue colors.}
  \label{fig:BR}
\end{center}\end{figure}

%
In Fig.~\ref{fig:BR}, we show the branching ratios of the Higgs boson $h$ into the new modes $h \to \phi \phi$ and $h \to Z'Z'$ as a function of $\sin \alpha$ by the solid and dashed curves respectively, where we fix the $Z'$ and $\phi$ masses to be  20 (0.02) GeV and 50 (0.05) GeV for the left (right) plot. The value of $g_X$ is taken to be 0.1, 0.01 and 0.001 for each plot as indicated by the black, red and blue curves.
We see that the branching ratios of 
$h \to \phi\phi$ and $h \to Z'Z'$ are going to be similar values 
for smaller $\sin\alpha$ and/or smaller masses of $\phi$ and $Z'$. 
This behavior can be understood by the following approximate formulae:  
\begin{align}
\begin{split}
& \Gamma (h \to Z' Z') = \frac{m_h^3 \alpha^2}{32 \pi v_\Phi^2} + \mathcal{O}(\alpha^4), \\
& \Gamma(h \to \phi \phi) = \frac{m_h^3\alpha^2}{32 \pi v_\Phi^2} \left(1 + \frac{2 m_\phi^2}{m_h^2} \right)^2 \left(1 + 2\alpha  \frac{v_\Phi}{v}\right) + \mathcal{O}(\alpha^4).
\end{split}
\label{eq:width-limit}
\end{align}
Thus, the above decay rates become the same if we neglect the 
$m_\phi^2/m_h^2$ term and the ${\cal O}(\alpha^3)$ contribution.
This can also be understood by noticing that 
the $h \to Z'Z'$ mode becomes one with the corresponding Goldstone boson in the final state, i.e., $h \to G_\Phi G_\Phi$
for $m_{Z'}/m_h \ll 1$ due to the equivalence theorem. 
In the limit of $\alpha \ll 1$ and $m_\phi/m_h \ll 1$, 
we obtain the relation among the couplings $\lambda_{\phi\phi h} = \lambda_{G_\Phi G_\Phi h} = -v \lambda_{H\Phi}/2 (=-\alpha m_h^2/2 v_\Phi)$, so that these two decay rates coincide with each other.  
It is also seen in Fig.~\ref{fig:BR} that 
smaller masses of $Z'$ and $\phi$ tend to give larger branching ratios of $h \to Z'Z'$ and $h \to \phi\phi$ for fixed values of $\sin\alpha$ and $g_X^{}$. 
This is because the smaller masses correspond to the smaller VEV $v_\Phi^{}$ from the relation $m_{Z'} \simeq g_X^{}v_\Phi^{}$, and the decay rates of these modes are proportional to $1/v_\Phi^2$ as seen in Eq.~(\ref{eq:width-limit}).



\section{One-loop corrections to the $hhh$ vertex \label{sec:hhh}}

We discuss the one-loop corrections to the Higgs trilinear vertex $hhh$. 
In order to obtain the finite results, we perform the renormalization of parameters, particularly those of the Higgs potential.  

We first shift the relevant bare parameters into the renormalized ones and the counterterms as follows:  
\begin{align}
T_\varphi \to \delta T_\varphi,~~  
m_\varphi^2 \to m_\varphi^2 + \delta m_\varphi^2,~~  
v \to v + \delta v,~~
v_\Phi^{} \to v_\Phi^{} + \delta v_\Phi^{},~~
\alpha \to \alpha + \delta \alpha + \delta \alpha_{\rm PT}^{}, \label{eq:shift1}
\end{align}
where $\varphi = h$ or $\phi$. 
For the shift of the mixing angle $\alpha$, we introduce not only the ordinal counterterm $\delta \alpha$ but also the one coming from the pinch-term~\cite{Cornwall:1989gv,Papavassiliou:1989zd} $\delta \alpha_{\rm PT}^{}$ which is required to maintain the gauge independent result~\cite{Krause:2016oke,Kanemura:2017wtm}\footnote{It has been known that the on-shell renormalization scheme, that is applied in this paper, generally gives rise to a gauge dependence in the renormalization of mixing parameter~\cite{Nielsen:1975fs}.  }. 
In Eq.~(\ref{eq:shift1}), $\delta T_\varphi$ represent the counterterms for the tadpole of $\varphi$. 
There are two ways for the treatment of $\delta T_\varphi$, i.e., the so-called standard tadpole scheme~\cite{Bohm:1986rj} and the alternative tadpole scheme~\cite{Fleischer:1980ub}. 
In the former scheme, tadpole counterterms are determined by requiring that renormalized scalar one-point functions vanish. 
On the other hand, in the latter, tadpole counterterms are not introduced, but all contributions from one-particle irreducible (1PI) diagrams are modified to include those with tadpole-inserted diagrams.
We analytically confirm that there is no difference between the renormalized $hhh$ vertex obtained by using the standard tadpole scheme and that by using the alternative tadpole scheme. 
In the following, the alternative tadpole scheme is applied to clarify the discussion.  
In addition to the parameter shift given in Eq.~(\ref{eq:shift1}), the bare fields are shifted as 
\begin{align}
\begin{pmatrix}
h \\
\phi
\end{pmatrix} \to 
\begin{pmatrix}
1 + \frac{1}{2}\delta Z_h & \delta Z_{h\phi} \\
\delta Z_{\phi h} & 1 + \frac{1}{2}\delta Z_\phi
\end{pmatrix}
\begin{pmatrix}
h \\
\phi
\end{pmatrix},  
\end{align}
where $\delta Z_{\phi h}$ and $\delta Z_{h \phi}$ can be decomposed as 
$\delta Z_{h \phi } = \delta \alpha + \delta C_h$ and $\delta Z_{\phi h} =  -\delta \alpha + \delta C_h$ with $\delta C_h$ being the wavefunction renormalization factor~\cite{Kanemura:2004mg}.

The renormalized scalar two-point functions are then expressed as 
\begin{align}
\hat{\Gamma}_{\varphi\varphi}(p^2) & = (p^2 -m_\varphi^2)(1 + \delta Z_\varphi) - \delta m_\varphi^2 + \Gamma_{\varphi\varphi}^{\rm 1PI}(p^2),~~(\varphi = h,~\phi) \\
\hat{\Gamma}_{\phi h}(p^2) & = (p^2 -m_h^2)\delta Z_{h\phi} + (p^2 - m_\phi^2)\delta Z_{\phi h} + \Gamma_{\phi h}^{\rm 1PI}(p^2), 
\end{align}
where $\Gamma_{XY}^{\rm 1PI}$ denotes the contribution from  1PI diagrams to the two-point function for $X$ and $Y$ particles.  
We impose the on-shell conditions as
\begin{align}
&\Re [\hat{\Gamma}_{\varphi\varphi}(m_\varphi^2)] = 0,\quad  \Re\left[\frac{d}{dp^2}\hat{\Gamma}_{\varphi\varphi}(p^2)\right]_{p^2=m_\varphi^2} = 1,\quad \Re[\hat{\Gamma}_{\phi h}(m_h^2)] = \Re[\hat{\Gamma}_{\phi h}(m_\phi^2)] = 0. 
\end{align}
These lead to 
\begin{align}
\delta m_\varphi^2 &= \Re[\Gamma_{\varphi\varphi}^{\rm 1PI}(m_\varphi^2)], \quad
\delta Z_\varphi   = -\Re \left[\frac{d}{dp^2}\Gamma_{\varphi\varphi}^{\rm 1PI}(p^2)
\right]_{p^2 = m_\varphi^2}, \\
\delta Z_{\phi h} & = -\frac{1}{m_h^2 - m_\phi^2} \Re[\Gamma_{\phi h}^{\rm 1PI}(m_h^2)]  , \quad
\delta Z_{ h\phi} = \frac{1}{m_h^2 - m_\phi^2}\Re[\Gamma_{\phi h}^{\rm 1PI}(m_\phi^2)]. 
\end{align}
The unpinched counterterm for the mixing angle $\delta \alpha$ is then given by 
\begin{align}
\delta \alpha & = \frac{\delta Z_{h\phi} -\delta Z_{\phi h} }{2} = 
\frac{1}{m_\phi^2 - m_h^2}\Re\left[\Gamma_{\phi h}^{\rm 1PI}(m_h^2) + \Gamma_{\phi h}^{\rm 1PI}(m_\phi^2) \right]. 
\end{align}
The pinch-term part is given by~\cite{Kanemura:2017wtm} 
\begin{align}
\delta\alpha_{\rm PT}^{} =  \frac{1}{m_\phi^2 - m_h^2}\Re\left[\Gamma_{\phi h}^{\rm PT}(m_h^2) + \Gamma_{\phi h}^{\rm PT}(m_\phi^2) \right],
\end{align}
where
\begin{align}
\Gamma_{\phi h}^{\rm PT}(p^2) = \frac{g^2}{64\pi^2}\sin2\alpha (2p^2 - m_h^2 - m_\phi^2)\left[B_0(p^2;m_W,m_W) +  \frac{1}{2c_W^2}B_0(p^2;m_Z,m_Z)\right],  
\end{align}
with $B_0$ being the Passarino-Veltman's scalar $B$ function~\cite{Passarino:1978jh}. 

The counterterms for the VEVs $\delta v$ and $\delta v_\Phi^{}$ can be determined from the renormalization of the EW parameters~\cite{Bohm:1986rj} which can significantly be different from those in the SM if we introduce the non-zero $Z$-$Z^\prime$ mixing. Although the $Z$-$Z^\prime$ mixing can play an important role in the decay of $Z^\prime$ even for a small mixing angle, its impact on the radiative correction to the $hhh$ vertex is negligibly small. 
We thus neglect the effect of the $Z$-$Z^\prime$ mixing, i.e., $\chi \to 0$ or equivalently $\epsilon \to 0$, on the renormalization of the $hhh$ vertex. 
In this setup, the renormalization of the EW parameters can be performed in a similar way as in the SM. Since the $Z$-$Z^\prime$ mixing is neglected, the transverse part of the renormalized two-point function for $Z^\prime$ is simply expressed as 
\begin{align}
\hat{\Gamma}_{Z^\prime Z^\prime}(p^2) = \Gamma_{Z^\prime Z^\prime}(p^2)
+(p^2- m_{Z^\prime}^2)(1 + \delta Z_{Z^\prime} )+ \delta m_{Z^\prime}^2, 
\end{align}
where $\delta m_{Z'}^2$ and $\delta Z_{Z'}$
are respectively the counterterms for the mass and wavefunction of $Z'$. 
Imposing the on-shell conditions, i.e., 
\begin{align}
\Re[\hat{\Gamma}_{Z^\prime Z^\prime}(m_{Z^\prime}^2)] = 0,\quad 
\Re\left[\frac{d}{dp^2}\hat{\Gamma}_{Z^\prime Z^\prime}(p^2)\right]_{p^2 = m_{Z^\prime}^2} = 1, 
\end{align}
we obtain
\begin{align}
\delta m_{Z^\prime}^2 = \Re[\Gamma^{\rm 1PI}_{Z^\prime Z^\prime}(m_{Z^\prime}^2)],\quad 
\delta Z_{Z^\prime} = -\Re\left[\frac{d}{dp^2}\Gamma^{\rm 1PI}_{Z^\prime Z^\prime}(p^2)\right]_{p^2=m_{Z^\prime}^2}.
\end{align}
%
The counterterm of the singlet VEV $\delta v_\Phi$ is then determined via the tree level relation, $v_\Phi^{} = m_{Z^\prime}^{}/g_X^{}$:
\begin{align}
\frac{\delta v_\Phi}{v_\Phi} 
& = \frac{\delta m_{Z'}^2}{2m_{Z'}^2} - \frac{\delta g_X^{}}{g_X^{}}
 = \frac{\Gamma^{\rm 1PI}_{Z'Z'}(m_{Z^\prime}^2)}{2m_{Z'}^2} - \frac{\delta g_X^{}}{g_X^{}}. \label{eq:delvp}
\end{align}
The counterterm for the $U(1)'$ gauge coupling $\delta g_X^{}$ is determined via the renormalization of the $\bar{f}fZ_\mu^\prime$ vertex as follows 
\begin{align}
\frac{\delta g_X^{}}{g_X^{}}  = \frac{1}{2}\frac{d}{dp^2}\Gamma_{Z'Z'}(p^2)\Big|_{p^2 = 0}. 
\end{align}
See Appendix~\ref{sec:ren_zp} for details of the renormalization of the $\bar{f}fZ'$ vertex. 

Now, we determine all the necessary counterterms for the computation of the renormalized $hhh$ vertex which is expressed as 
\begin{align}
\hat{\Gamma}_{hhh} & = \Gamma_{hhh}^{\rm tree}  + \Gamma_{hhh}^{\rm 1PI} + \delta\Gamma_{hhh}, 
\end{align}
where the first, second and third terms respectively denote the contributions from the tree level diagram, 1PI diagrams and the counterterms.  
The tree-level contribution is given by 
\begin{align}
\Gamma_{hhh}^{\rm tree} = 3! \lambda_{hhh},     
\end{align}
where $\lambda_{hhh}$ is given in Eq.~(\ref{eq:hhh}). 
The counterterm contribution is given by 
\begin{align}
\delta\Gamma_{hhh} & = 3!\left(\delta \lambda_{hhh} + \frac{3}{2}\lambda_{hhh}\delta Z_h + \lambda_{\phi hh}\delta Z_{\phi h} \right), 
\end{align}
where 
\begin{align}
\delta \lambda_{hhh}  &= \frac{m_h^2\cos^3\alpha}{2v}\frac{\delta v}{v} + \frac{m_h^2\sin^3\alpha}{2v_\Phi}\frac{\delta v_\Phi}{v_\Phi} - \frac{1}{2vv_\Phi}(v\sin^3\alpha + v_\Phi \cos^3\alpha)\delta m_h^2\notag\\
&+ \frac{3m_h^2}{4vv_\Phi}\sin2\alpha (v_\Phi \cos\alpha - v\sin\alpha)(\delta\alpha + \delta\alpha_{\rm PT}). 
\end{align}

We will see in the next section that the deviation of the $hhh$ coupling from the SM prediction is very small such as $\sim 1\%$ or smaller.   
This is because the contribution to the $hhh$ vertex from the $Z'$ loop is suppressed by $(g_X^{2} \sin\alpha)^3$ which is negligibly small for $g_X^{} \ll 1$. 
In addition, the contribution from the $\phi$ loop is 
proportional to $\lambda_{\phi\phi h}^3$ which is suppressed by $\sin^3\alpha$ for $|\sin\alpha| \ll 1$. 
Thus, the situation is completely different from models with singlet-scalar extensions without a $U(1)'$ symmetry, where the latter can predict ${\cal O}(100)$ \% deviation from the SM, see e.g.,~\cite{Kanemura:2016lkz}. This can be realized because more parameters appear in the potential and the $\lambda_{\phi\phi h}$ coupling can independently be taken of the mixing angle. 

\section{Phenomenology\label{sec:pheno}}

In this section, we discuss multi-$Z'$ signatures via the decays of $h$ and $\phi$ in our models with the spontaneously broken $U(1)'$ symmetry at the LHC.
There are five new independent parameters in our scenario, which are chosen to be $\{m_\phi,m_{Z'},\sin\alpha,g_X^{},\epsilon\}$. 
For simplicity, we take the kinetic mixing parameter $\epsilon$ to be zero except for the $U(1)_D$ scenario. In the latter, the phenomenology does not depend on the value of $\epsilon$ as far as $\epsilon \ll 1$\footnote{For instance, we can consider the case with $e\epsilon \sim 10^{-4}$ which is allowed by current experimental data~\cite{Bauer:2018onh}. } because we consider the $Z'$ production only via the Higgs decay and the branching ratio of $Z'$ is simply determined by the electric charge of SM fermions, see Sec.~\ref{subsec:gauge-sector}.   
For the masses of $\phi$ and $Z'$, we consider three reference sets that are (i) small mass case $\{m_{Z'}, m_\phi \} = \{0.02, 0.05\}$ GeV, (ii) middle mass case $\{m_{Z'}, m_\phi \} = \{20, 50\}$ GeV, 
and (iii) large mass case $\{m_{Z'}, m_\phi \} = \{200, 800\}$ GeV, all of which  
the $\phi \to Z'Z'$ mode is kinematically allowed.
We then discuss the phenomenology on the parameter space of $\{g_X, \sin \alpha\}$.

\subsection{Small mass case: $\{m_{Z'}, m_\phi \} = \{0.02, 0.05\}$ GeV} 

 \begin{figure}[tb]
 \begin{center}
\includegraphics[width=5cm]{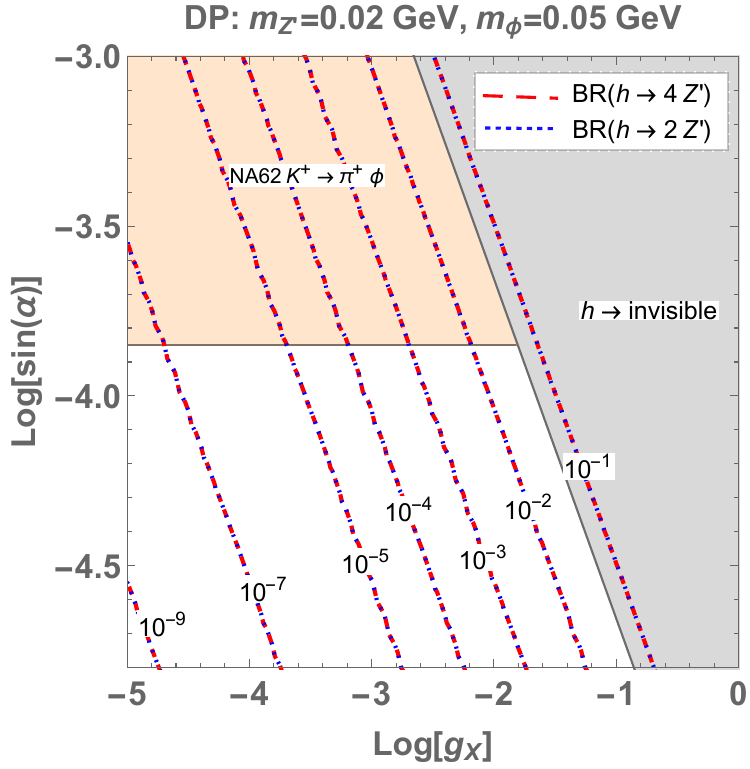}  \
\includegraphics[width=5cm]{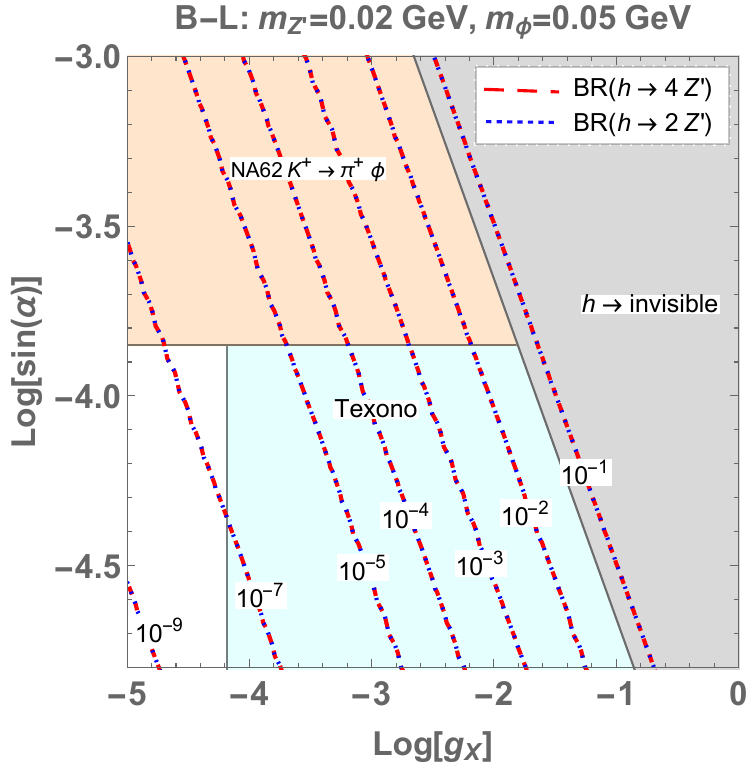} \\
\includegraphics[width=5cm]{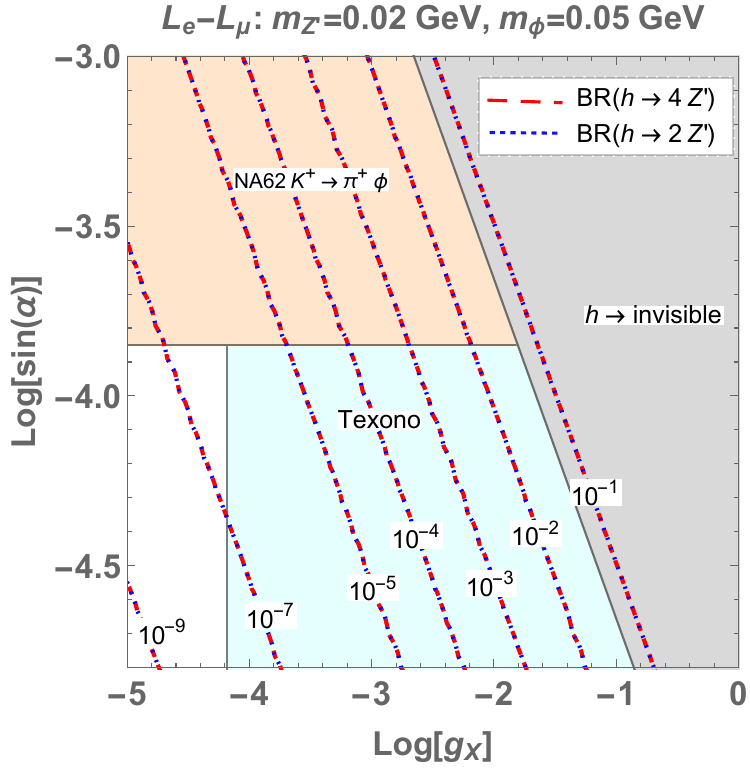} \
 \includegraphics[width=5cm]{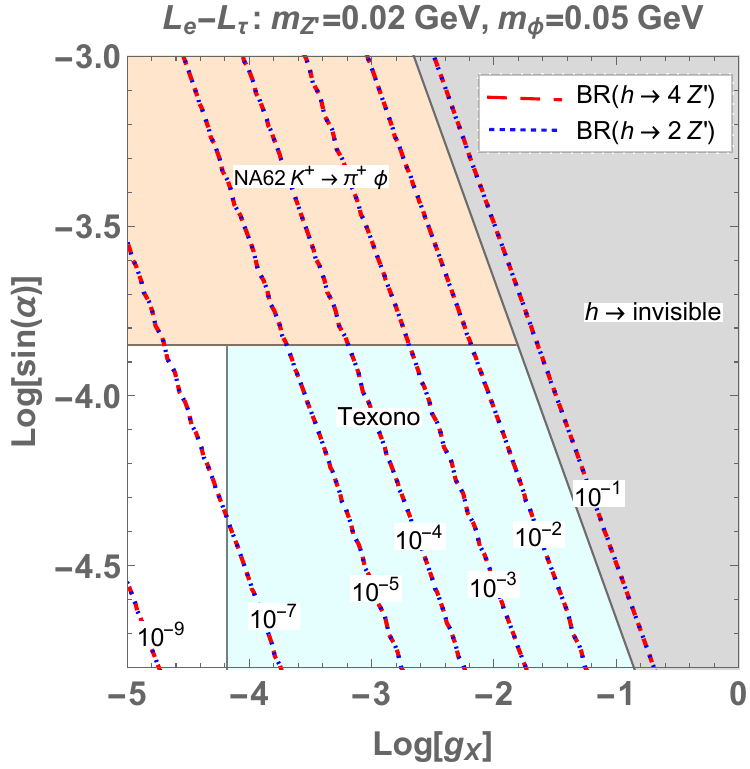}  \
\includegraphics[width=5cm]{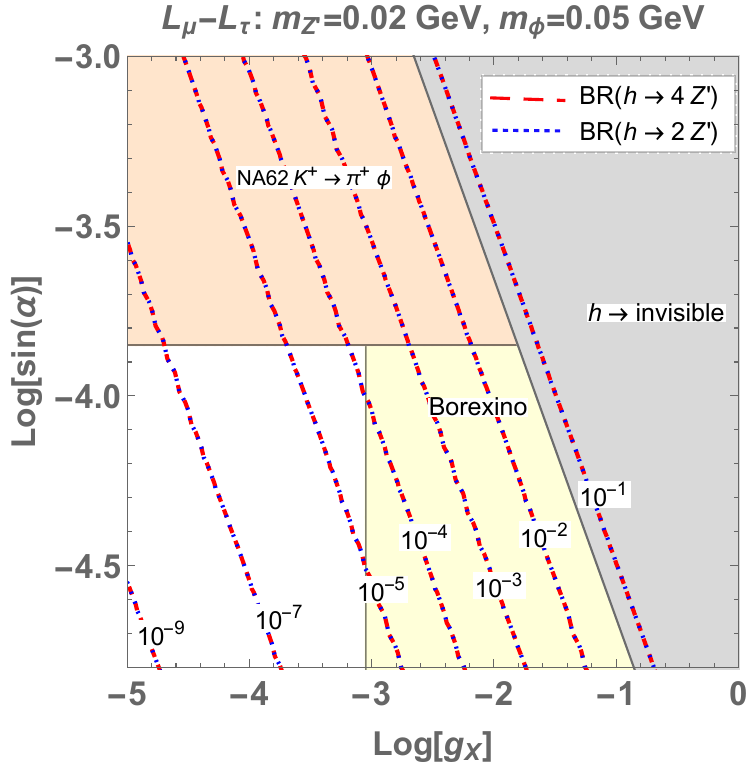} 
\caption{Contour plots of the branching ratios BR$(h \to 4 Z')$ and BR$(h \to 2 Z')$ in the $U(1)'$ models where DP stands for dark photon, i.e., the $U(1)_D$ model. The 
gray region labeled by $h \to $invisible is excluded by the search for $h \to $invisible signal at the LHC. 
The orange region is excluded by $K^+ \to \pi^+ \phi$ search at NA62. }
\label{fig:low-mass}
\end{center}
\end{figure}

In this case, we focus on the $h$ decay into $Z'Z'$ and $4Z'$ final states and estimate the branching fraction of these modes.
In principle, the singlet-like scalar $\phi$ can  be produced via the gluon fusion process, but its decay products have too small energies to be detected at LHC. We thus do not consider the gluon fusion production of $\phi$. 
On the other hand, we need to take into account the production of $\phi$ via meson decays through the scalar mixing effect. 
In particular, we impose the strongest bound on the scalar mixing angle coming from the $K^+ \to \pi^+ \phi$ search at NA62~\cite{NA62:2021zjw}.
Constraints from the discovered Higgs boson decay should also be considered. 
Here, we note that 2 or 4 SM fermion pairs produced from the decay chains of $h \to Z' Z' \to (f \bar f)(f \bar f) $ and $h \to \phi \phi \to (Z' Z')(Z' Z') \to  (f \bar f f \bar f)(f \bar f f \bar f)$ are 
highly collimated due to the lightness of $\phi$ and $Z'$. 
Thus, typical constraints from multi-lepton final states cannot be applied directly.
Instead, we consider the bounds from ``$h \to {\rm invisible}$"~\cite{CMS:2022qva,ATLAS:2023tkt} and ``$h \to {\rm BSM}$"~\cite{ATLAS:2019nkf} where ``BSM" is any BSM final states.
Among them the strongest bound is BR$(h \to {\rm invisible}) < 0.107$~\cite{ATLAS:2023tkt}, so that  we impose the following constraint as a conservative limit 
\begin{equation}
{\rm BR}(h \to Z' Z') + {\rm BR}(h \to \phi \phi) < 0.107. 
\end{equation}

In addition, we impose constraints on the new gauge coupling $g_X$ from various experiments that can test $Z'$ interaction with SM fermions.  
For the $Z'$ coupling, the strongest constraints are obtained by the Texono experiment~\cite{TEXONO:2009knm} in the $U(1)_{B-L}$, 
$U(1)_{L_e - L_\mu}$ and 
$U(1)_{L_e-L_\tau}$ models and by the Borexino experiment~\cite{Bellini:2011rx} in the  $U(1)_{L_\mu - L_\tau}$ model, which test interactions between $Z'$ and neutrinos. Bounds on the $Z'$ coupling and mass are comprehensively summarized in Refs.~\cite{Ilten:2018crw, Bauer:2018onh, Asai:2022zxw, Asai:2023mzl, KA:2023dyz}. 

In Fig.~\ref{fig:low-mass}, we show the contour plots for BR$(h \to Z' Z')$ and BR$(h \to 4Z')$ on the $\{g_X, \sin \alpha\}$ plane in each model.
The gray region is excluded by the constraint from the branching ratio of $h \to ``{\rm invisible}"$.
The light cyan regions for the $U(1)_{B-L}$, $U(1)_{L_e-L_\mu}$ and $U(1)_{L_e - L_\tau}$ cases are excluded by the Texono experiment.
The light yellow region for the $U(1)_{L_\mu - L_\tau}$ case is excluded by the Borexino experiment.
In addition, the light orange region is excluded by the $K^+ \to \pi^+ \phi$ search at NA62.
We see that $\sin \alpha$ is highly restricted to be smaller than $\sim 10^{-4}$
due to the constraints from $K^+ \to \pi^+\phi$ and BR$(h \to Z' Z')\simeq \ $BR$(h \to \phi \phi \to 4Z')$ as we discussed below Eq.~\eqref{eq:width-limit}.
It is thus found that sizable branching ratios cannot be obtained when SM fermions are charged under $U(1)'$. 
On the other hand, the branching ratio of BR$(h \to Z'Z'/4Z') \lesssim 0.05$ can be realized for the $U(1)_D$ case since the
new gauge coupling is not directly constrained from the experiments which have a sensitivity to the $Z'f\bar{f}$ interactions.
Therefore, it is possible to have a large number of events at the LHC for the $U(1)_D$  case, where $Z'$ decays into $e^+e^-$ pair via the kinetic mixing with almost $100 \%$. Thus, the signal processes are 
\begin{equation}
h \to Z'Z' \to (e^+e^-) (e^+ e^-), \quad h \to \phi \phi \to  (Z' Z') (Z' Z') \to (e^+ e^- e^+ e^-) (e^+ e^- e^+ e^-),
\end{equation}
where the particles inside the parentheses are highly collimated. 
Thus, these signals can be seen as electron-jets where we need dedicated simulation studies to clarify their detectability at collider experiments. 
We can also have $\mathcal{O}(10^{-4})$ for BR$(h \to Z'Z'/4Z')$ in the $U(1)_{L_\mu - L_\tau}$ case but $Z'$ only decays into neutrinos as long as we take the kinetic mixing to be zero while $Z' \to e^+ e^-$ is possible if a non-zero kinetic mixing is taken as in the $U(1)_D$ case.
For the $U(1)_{L_e - L_{\mu}}$ and $U(1)_{L_e - L_\tau}$ cases, we can have BR$(h \to Z'Z'/4Z') \lesssim 10^{-6}$ while BR$(Z' \to e^+e^-)$ is given to be $\sim 50\%$.
Therefore, the $U(1)_D$ case is the most promising in this mass scale to search for the signal at the LHC.

\subsection{Middle mass case: $\{m_{Z'}, m_\phi \} = \{20, 50\}$ GeV}

\begin{table}[t]
  \begin{center}
    \begin{tabular}{|c|c|c|c|c|c|c|}\hline
& $~e^+e^-~$ & $~\mu^+\mu^-~$ & $~\tau^+ \tau^-~$ & $~\nu \bar \nu~$ & $~jj~$ & $~b \bar b~$ \\ \hline
$U(1)_D$ & 0.15 & 0.15 & 0.15 & 0 & 0.50 & 0.049 \\
$U(1)_{B-L}$ & 0.16 & 0.16 & 0.16 & 0.24 & 0.22 & 0.054 \\
$U(1)_{L_e - L_\mu}$ & 0.33 & 0.33 & 0 & 0.33 & 0 & 0 \\
$U(1)_{L_e - L_\tau}$ & 0.33 & 0 & 0.33 & 0.33 & 0 & 0 \\
$U(1)_{L_\mu - L_\tau}$ & 0 & 0.33 & 0.33 & 0.33 & 0 & 0 \\ \hline
    \end{tabular}
  \end{center}
  \caption{The BRs of $Z' \to f \bar f$ decays for $m_{Z'} = 20$ GeV where kinetic mixing is ignored except for the $U(1)_D$ case.}
  \label{tab:BRs}
\end{table}

 \begin{figure}[tb]
 \begin{center}
\includegraphics[width=5cm]{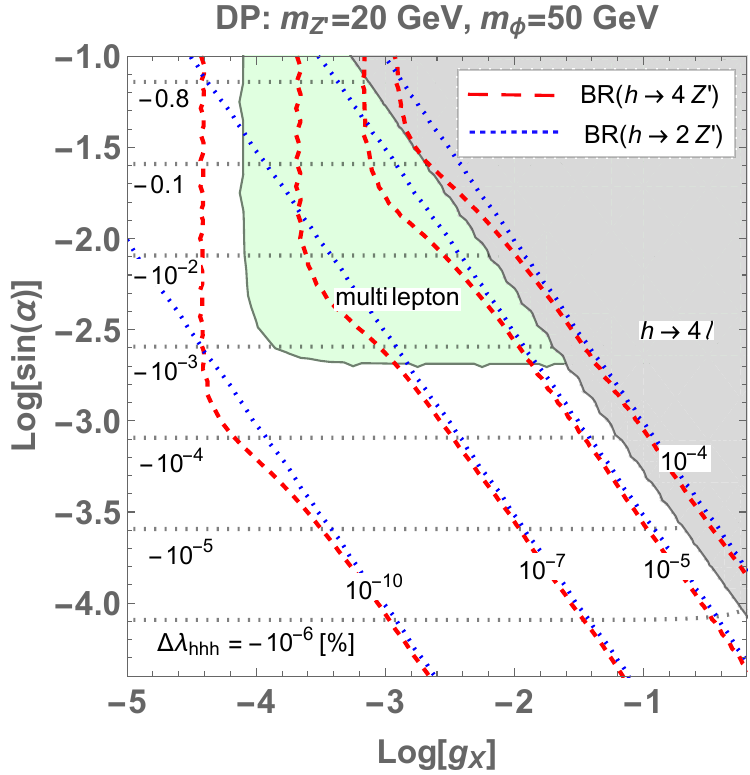}  \
\includegraphics[width=5cm]{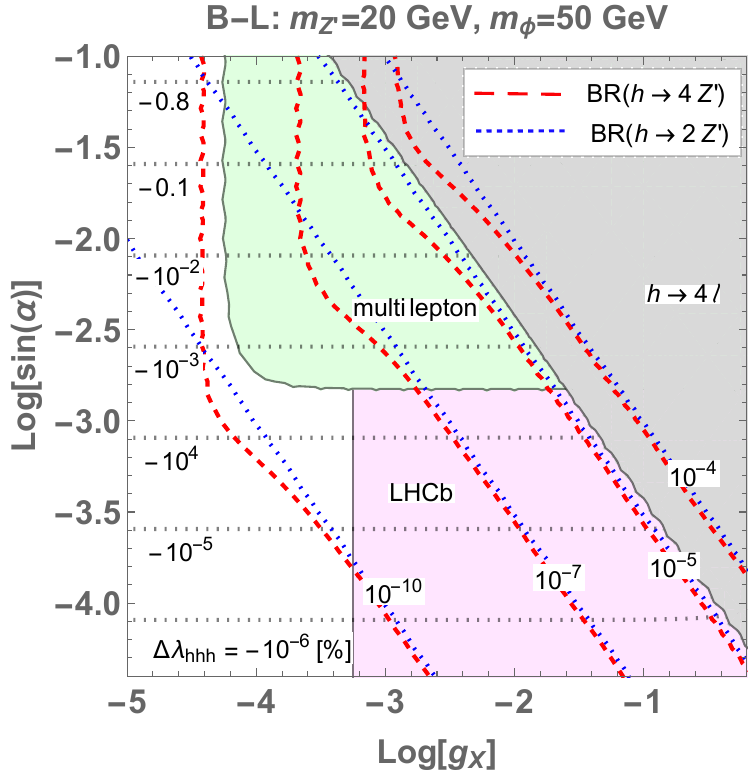} \\
\includegraphics[width=5cm]{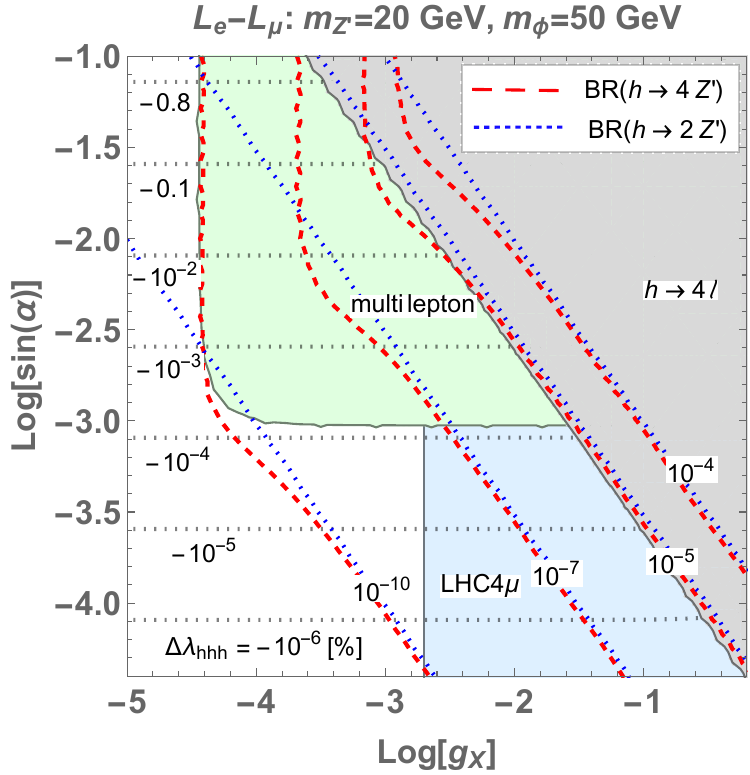} \
 \includegraphics[width=5cm]{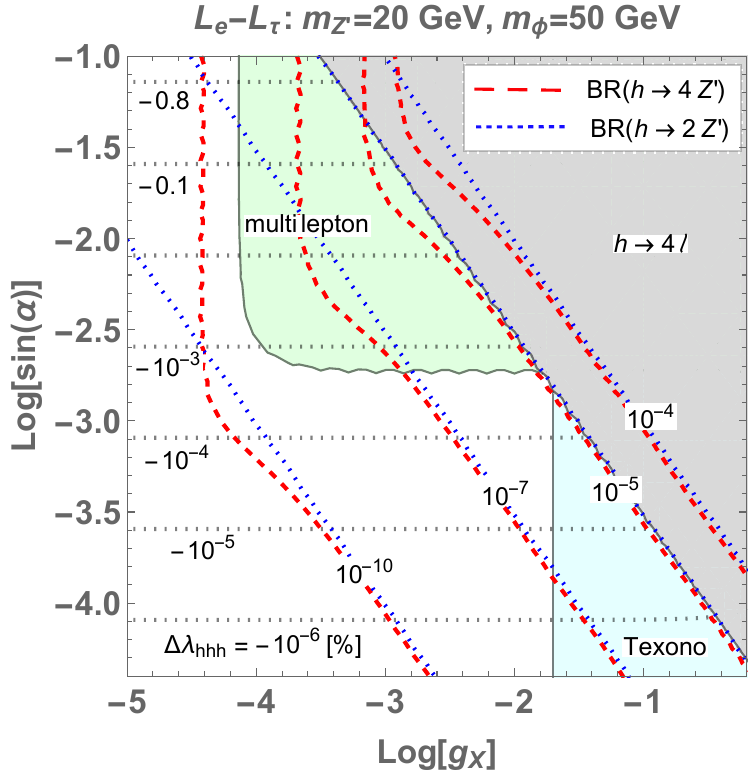}  \
\includegraphics[width=5cm]{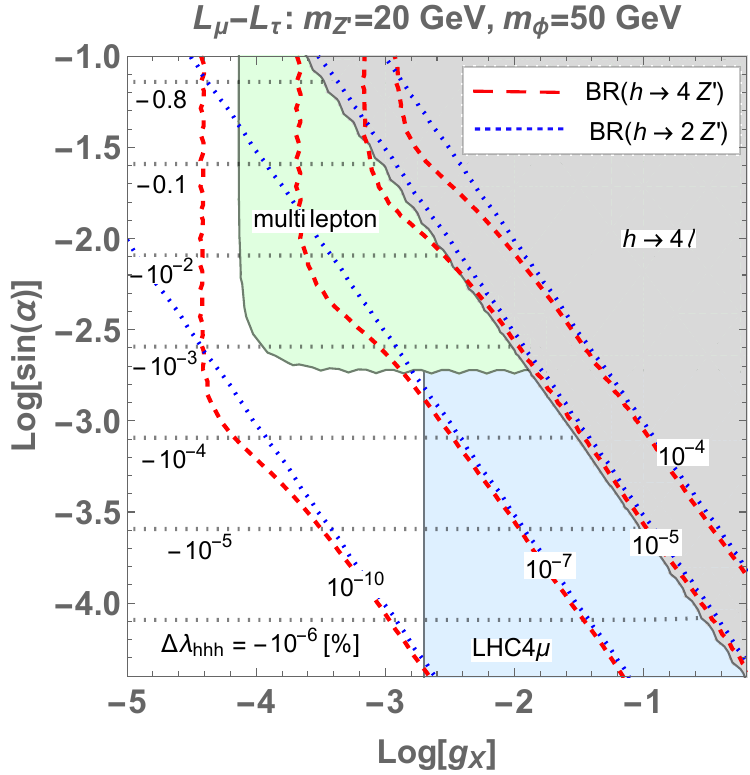} 
\caption{Contour plots of the branching ratios BR$(h \to 4 Z')$ and BR$(h \to 2 Z')$ in the  $U(1)'$ models. The gray region is excluded by the search for $h \to 4 \ell$ signals at the LHC. 
The light green region is excluded by the multilepton search at the LHC where leptons come from $\phi \to Z' Z' \to 4 \ell$ decay after $\phi$ production via gluon fusion. The light magenta region is excluded by LHCb measurement for the $U(1)_{B-L}$ case. The light blue regions are excluded by LHC $\mu^+ \mu^- Z'(\to \mu^+ \mu^-)$ searches for $U(1)_{L_{e(\mu)-L_{\mu(\tau)}}}$ case. The light cyan region is excluded by Texono experiment for $U(1)_{L_e - L_\tau}$. The dashed contours indicate values of deviation of the $hhh$ coupling from the SM at one-loop level.  }
\label{fig:middle-mass}
\end{center}
\end{figure}

As in the small mass case, we also focus on the new decay modes of the Higgs boson $h$.
In this case, we consider the constraints from the $h \to (\ell^+ \ell^-)(\ell^+ \ell^-)$ searches~\cite{CMS:2021pcy,ATLAS:2021ldb} and the non-resonant multi-lepton searches~\cite{ATLAS:2021wob} at the LHC where the latter can be induced from the gluon fusion production of $\phi$ followed by the decay chain of $\phi \to Z'Z' \to (\ell^+ \ell^-) (\ell^+ \ell^-)$.  The cross section of $gg \to \phi$ is estimated by {\tt CalcHEP\_3.8}~\cite{Belyaev:2012qa} implementing the effective interaction for $\phi$-gluon-gluon, and we get $\sigma(gg \to \phi) \simeq \sin^2 \alpha \times 250$ pb applying K-factor 1.6~\cite{Djouadi:2005gi} for $m_\phi =$ 50 GeV and $\sqrt{s} = 13$ TeV case.
Here, we use the branching ratios of $Z'$ given in Table~\ref{tab:BRs} for the estimation of the bounds.
In addition, we apply to the constraints on the gauge coupling $g_X^{}$ from $Z'$ searches where the strongest constraint comes from LHCb~\cite{LHCb:2019vmc} for $U(1)_{B-L}$, $pp \to \mu^+ \mu^- Z'(\to \mu^+ \mu^-)$ searches at the LHC~\cite{CMS:2018yxg,ATLAS:2023vxg} for $U(1)_{L_{e(\mu)}- L_{\mu(\tau)}}$ and neutrino scattering measurement at Texono~\cite{TEXONO:2009knm} for $U(1)_{L_e - L_\tau}$.

In Fig.~\ref{fig:middle-mass}, we show the contour plots for BR$(h \to Z' Z')$ and BR$(h \to 4Z')$ on the $\{g_X, \sin \alpha\}$ plane in each model.
The gray and green regions are excluded by the $h \to (\ell^+ \ell^-)(\ell^+ \ell^-)$ searches and the non-resonant multi-lepton searches, respectively. 
The light magenta region is excluded by the LHCb measurement for the $U(1)_{B-L}$ case. 
Also, the light blue regions are excluded by the $\mu^+ \mu^- Z'(\to \mu^+ \mu^-)$ searches at LHC for the $U(1)_{L_{e(\mu)}- L_{\mu(\tau)}}$ cases.
In addition, the light cyan region is excluded by the Texono experiment for the $U(1)_{L_e - L_\tau}$ case.
We find that BR$(h \to Z'Z') \simeq \text{BR}(h \to 4Z') \lesssim 3 \times 10^{-5}$ is allowed for the $U(1)_D$ case and the value of $\lesssim 10^{-5}$ is allowed for the $L_e - L_\tau$ case.
On the other hand, the allowed value of the branching ratio is much smaller for the $U(1)_{B-L}$ and $U(1)_{L_{e(\mu)}- L_{\mu(\tau)}}$ cases than that of the $U(1)_D$ or $U(1)_{L_e-L_\tau}$ case due to the stronger constraint from LHCb and $\mu^+\mu^- Z'$ searches.
From the maximal branching ratio of $\mathcal{O}(10^{-5})$, we expect the cross section of the $gg \to h \to Z'Z'/4Z'$ processes to be 
${\cal O}(1)$ fb which is obtained by $\sigma_{gg \to h} \sim 50$ pb for the collision energy of 13 and 14 TeV. 
The signals from the multi-$Z'$ production depend on $U(1)'$ via branching ratios of $Z'$ as shown in Table~\ref{tab:BRs}.
In addition, we show the deviation of the $hhh$ coupling from the SM at one-loop level, denoting $\Delta \lambda_{hhh}$, by partially modifying the {\tt H-COUP} package~\cite{Kanemura:2017gbi,Kanemura:2019slf,Aiko:2023xui}. 
We find that the deviation is almost independent of $g_X$ except for the region with $g_X = \mathcal{O}(1)$, and the magnitude of the deviation can be $\mathcal{O}(1) \%$ for $\sin \alpha = \mathcal{O}(0.1)$.

\subsection{Large mass case: $\{m_{Z'}, m_\phi \} = \{200, 800\}$ GeV}

 \begin{figure}[tb]
 \begin{center}
\includegraphics[width=5cm]{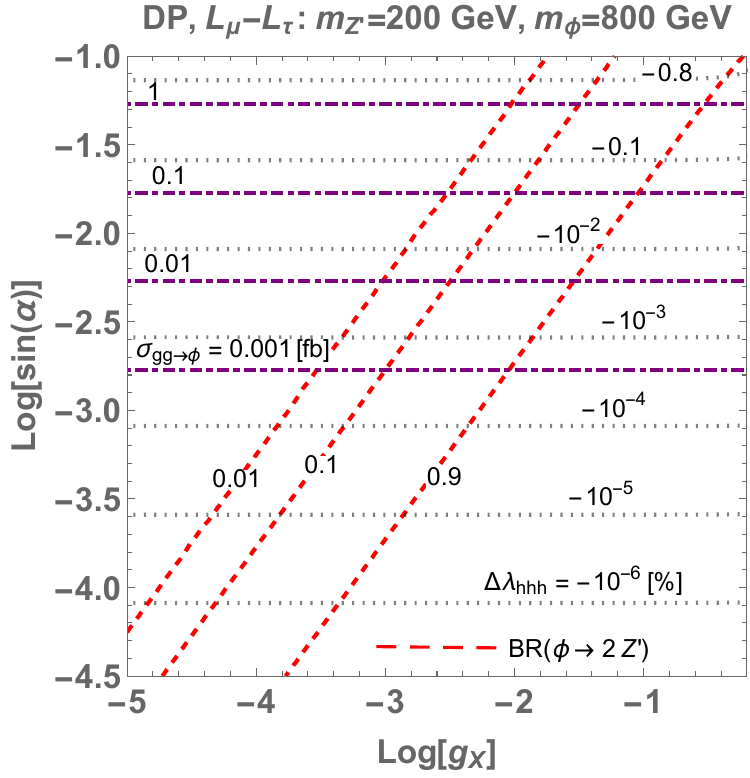}  \quad 
\includegraphics[width=5cm]{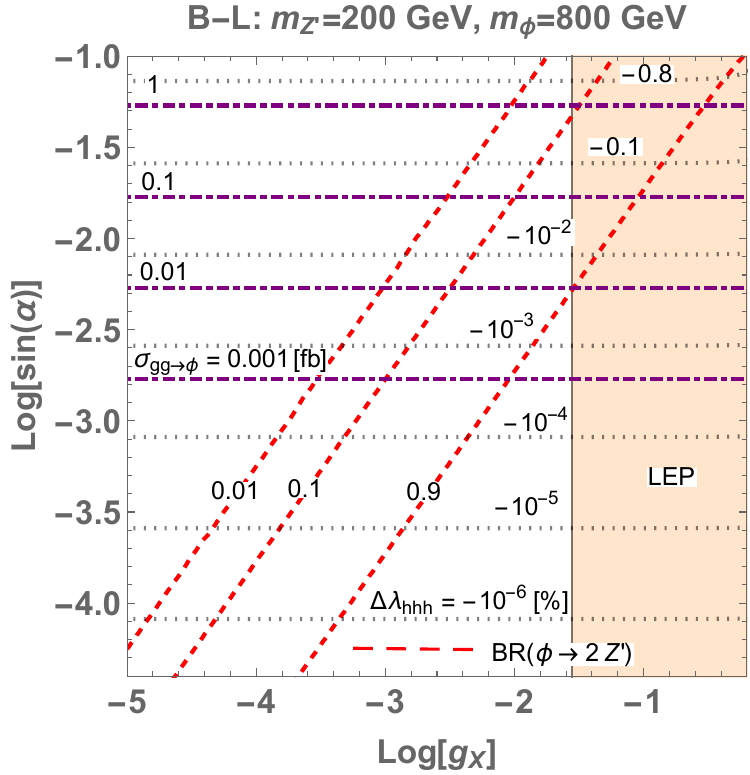} \quad
\includegraphics[width=5cm]{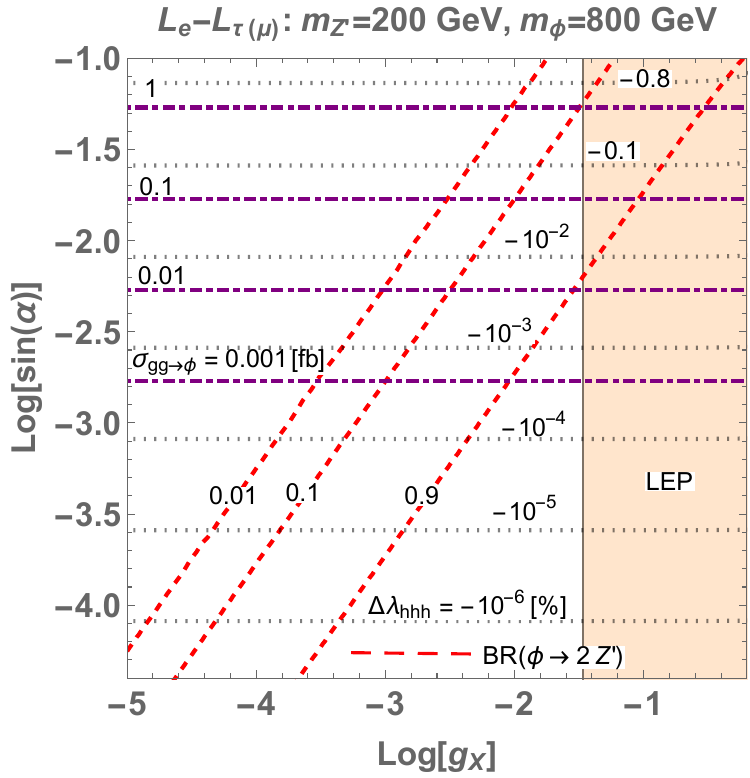} \\
\caption{Contours of the branching ratios BR$(\phi \to 2 Z')$ are shown by the red dashed curves. The purple dash-dotted contours represent $\phi$ production cross section via gluon fusion with $\sqrt{s} = 13$ TeV. The light orange region is excluded by the LEP experiment. The dashed contours indicate values of deviation of the $hhh$ coupling from the SM at one-loop level.   }
\label{fig:high-mass}
\end{center}
\end{figure}

In this case, the Higgs boson $h$ cannot decay into $\phi \phi$ or $Z'Z'$.
We thus consider the production and decay of $\phi$. 
By using {\tt CalcHEP\_3.8}, we obtain 
$\sigma(gg \to \phi) \simeq \sin^2 \alpha \times 350$ fb for $m_\phi =$ 800 GeV and $\sqrt{s} = 13$ TeV.

In Fig.~\ref{fig:high-mass}, BR$(\phi \to Z' Z')$ is shown as the red dashed lines on the $\{g_X, \sin \alpha\}$ plane.  
The lower limit on $m_{Z'}/g_X$ is given by the LEP experiment as $m_{Z'}/g_X > 6.9 \, (5.8)$ TeV for the $U(1)_{B-L}$ ($U(1)_{L_e -L_\mu}$ and $U(1)_{L_e -L_\tau}$) case~\cite{ALEPH:2013dgf}, which is translated into the upper limit on $g_X$ shown by the light orange color in the figure.
We also show the values of the gluon fusion cross section by the purple dot-dashed horizontal lines.
It is found that the cross section for 
the $gg \to \phi \to Z'Z'$ process can maximally be ${\cal O}(1)$ fb and ${\cal O}(0.1)$ fb for the $U(1)_D$ and $U(1)_{L_\mu - L_\tau}$ cases and the $U(1)_{B-L}$, $U(1)_{L_e - L_\mu}$ and $U(1)_{L_e - L_\tau}$ cases, respectively. 
We note that the decay branching ratios of $Z'$ into the SM fermions are almost the same as those shown in Table~\ref{tab:BRs}.
In addition, we show the deviation in the $hhh$ coupling from the SM which has almost the same behavior as the middle mass case in Fig.~\ref{fig:middle-mass}.

In Table~\ref{tab:4}, we summarize the approximate upper bounds on BR$(h \to Z'Z'/\phi \phi)$ and the signatures expected from the multi-$Z'$ production in each model.
For each mass case, the $U(1)_D$ model can provide the largest number of events due to the weaker constraint on the gauge coupling $g_X$, in which $Z'$ does not decay into neutrinos so that the final states of the signature do not include sizable missing transverse energy unlike all the other $U(1)'$ models.  
In each model, we expect the final state with multi-jets and/or multi-leptons for the middle and large mass cases,  where the fractions of jets and leptons 
depend on the model via the branching ratio of $Z'$ (see Table.~\ref{tab:BRs} for example). 
Thus, detailed analyses of the multi-$Z'$ signature can determine the $U(1)'$ model. 
Further simulation studies are beyond the scope of this work, and we leave them as future works.

\begin{table}[t]
\begin{tabular}{c||c c c c c c }\hline  
\multicolumn{6}{c}{(i) Small mass case } \\ \hline
&  ~$U(1)_D$~ & ~$U(1)_{B-L}$~ & ~$U(1)_{L_e - L_\mu}$~ & ~$U(1)_{L_e - L_\tau}$~ & ~$U(1)_{L_\mu - L_\tau}$~   \\\hline
BR$(h \to Z'Z'/ \phi \phi)$  & $\lesssim 0.05$  & $\lesssim 10^{-6}$  & $\lesssim 10^{-6}$ & $\lesssim 10^{-6}$ & $\lesssim 10^{-4}$      \\
Signal of Multi-$Z'$  &  ~e-jets~ & ~e-jets (+$\slashed{E}_T$)~ & ~e-jets ($+\slashed{E}_T$)~ & ~e-jets ($+\slashed{E}_T$)~ & ~$\slashed{E}_T$ [+e-jets]~  \\\hline 
\multicolumn{6}{c}{(ii) Middle mass case } \\ \hline
&  ~$U(1)_D$~ & ~$U(1)_{B-L}$~ & ~$U(1)_{L_e - L_\mu}$~ & ~$U(1)_{L_e - L_\tau}$~ & ~$U(1)_{L_\mu - L_\tau}$~   \\\hline
BR$(h \to Z'Z'/\phi\phi)$  & $\lesssim 10^{-5}$  & $\lesssim 10^{-7}$  & $\lesssim 10^{-5}$ & $\lesssim 10^{-7}$ & $\lesssim 10^{-7}$      \\
Signal of Multi-$Z'$   & ~multi-$j/\ell$~ & ~multi-$j/\ell$ ($+\slashed{E}_T$)~ & ~multi-$\ell$ ($+\slashed{E}_T$)~ & ~multi-$\ell$ ($+\slashed{E}_T$)~ & ~multi-$\ell$ ($+\slashed{E}_T$)~  \\\hline 
\multicolumn{6}{c}{(iii) Large mass case } \\ \hline
Signal of Multi-$Z'$   & ~multi-$j/\ell$~ & ~multi-$j/\ell$ ($+\slashed{E}_T$)~ & ~multi-$\ell$ ($+\slashed{E}_T$)~ & ~multi-$\ell$ ($+\slashed{E}_T$)~ & ~multi-$\ell$ ($+\slashed{E}_T$)~  \\\hline 
\end{tabular}
\caption{Summary of upper bounds of BR$(h \to Z'Z'/\phi \phi)$ [No upper bound of BR$(\phi \to Z'Z')$ for case (iii)] and signatures in each model where e-jets stand for ``electron-jet", $\slashed{E}_T$ is missing transverse energy, and multi-$\ell/j$ denotes multi-leptons/jets signal. The e-jets for $U(1)_{L_\mu -L_\tau}$ in case (i) can be obtained via kinetic mixing effect.}\label{tab:4}
\end{table}

\section{Summary and discussion\label{sec:summary}}

In this paper, we have explored models with a spontaneously broken $U(1)'$ gauge symmetry, and have shown possibilities for obtaining multi-$Z'$ signatures
from decays of the Higgs bosons. 
In particular, we have focused on anomaly-free $U(1)'$ models without introducing new fermions (except for right-handed neutrinos), in which the SM Higgs field is not charged under the symmetry. 
More specifically, we have considered the models with the $U(1)_{B-L}$, $U(1)_{L_i - L_j}$ and hidden $U(1)_D$ symmetries where the last one is also known as the dark photon model.
%
The $U(1)'$ symmetry is spontaneously broken by the VEV of a new scalar field $\Phi$ which is singlet under the SM gauge symmetry. 

We then have discussed decays of scalar bosons $h$ (SM-like) and $\phi$ (singlet-like) and a new gauge boson $Z'$. We have seen that the new decay channels $h \to Z'Z'$ and $h \to \phi\phi$ can be significant for a larger mixing angle $\sin\alpha$ and/or a larger new gauge coupling $g_X^{}$.  
Furthermore, we have evaluated one-loop corrections to the Higgs trilinear vertex by applying the on-shell renormalization scheme without gauge dependence. 

Finally, multi-$Z'$ signatures from the scalar boson decays have been discussed, where we have chosen three reference points for the masses of $\phi$ and $Z'$, i.e., $\{m_{Z'}, m_\phi \}$ to be (i) $\{0.02, 0.05\}$ GeV, (ii) $\{20, 50\}$ GeV, and (iii) $\{200, 800\}$ GeV. 
For case (i), we have shown the branching ratios of 
$h \to Z'Z'$ and $h \to \phi\phi \, (\phi\to Z'Z')$ modes 
taking into account constraints on the new gauge coupling and the scalar mixing from $h \to \text{invisible}$ decay, $K^+ \to \pi^+ \phi$ process and searches for $Z'$ interactions with SM fermions by various experiments.
For the dark photon case, it has been found that we can have the sizable branching ratios for $h \to Z'Z' \to (e^+ e^-) (e^+ e^-)$ and $h \to \phi \phi \to (Z' Z')(Z' Z') \to (e^+ e^- e^+ e^-)(e^+ e^- e^+ e^-)$ where particles inside bracket will be highly collimated, while those of $h \to \phi\phi/Z'Z'$ are much smaller in the other $U(1)'$ models than the dark photon model. Thus, we can expect ``electron-jets" signals from the processes at collider experiments.
For case (ii), we have also shown the Higgs decay branching ratios. 
In this case, we have discussed constraints from multi-lepton signals from the decays of $h$ and $\phi$ at LHC. 
The new gauge coupling is also constrained from searches for $Z'$ by flavor experiments.
It has been found that BR$(h \to 2 Z' )\simeq$BR$(h \to 2\phi \to 4Z') \sim \mathcal{O}(10^{-5})$ is allowed in the dark photon and $U(1)_{L_e-L_\tau}$ cases while the branching ratios have to be smaller by orders in the other cases. 
In these cases, we obtain the cross section of the $gg \to h \to Z'Z'/4Z'$ process to be ${\cal O}$(1-0.1) fb which can be detected by future LHC experiments. 
In addition, we have shown the deviation of $hhh$ coupling from the SM, and found that its magnitude is less than 1\%. 
For case (iii), the SM-like Higgs boson does not decay into $\phi$ or $Z'$, and we have discussed production and decays of $\phi$. In this mass range, $g_X^{}$ and $\sin\alpha$ are not strongly constrained by experiments compared with the other mass ranges. 
We have then shown the production cross section of $\phi$, BR$(\phi \to Z'Z')$ and the deviation of the  $hhh$ coupling.
In this case, $\phi$ production induces multi-fermion signals via $Z'$ where the fraction of signals depends on the $U(1)'$ type.

In summary, we have found possibilities for obtaining a detectable number of multi-$Z'$ events from the decay of $h$ and the production/decay of $\phi$, where a type of signals depends on the mass scale of new bosons as well as that of the $U(1)'$ symmetry.
Exploration of such signals is important to test spontaneously broken $U(1)'$ models. 
Detailed analysis of signals is required for further testing models and it will be done in upcoming works.

\section*{Acknowledgments}
The work was supported by the Fundamental Research Funds for the Central Universities (T.~N.). 

\begin{appendix}

\section{Renormalization of the $\bar{f}fZ_\mu^\prime$ vertex} \label{sec:ren_zp}

The renormalized $\bar{f}fZ^\prime_\mu$ vertex can be decomposed into the following form factors: 
\begin{align}
\hat{\Gamma}_{Z'ff}^\mu = g_X^{}\gamma^\mu
(\hat{\Gamma}_{Z'ff}^V -\gamma_5 \hat{\Gamma}_{Z'ff}^A) + i\sigma^{\mu\nu}q_\nu
\hat{\Gamma}_{Z'ff}^M, 
\end{align}
where $q^\mu$ is the outgoing four-momentum for $Z'$, and
\begin{align}
\sigma^{\mu\nu} = \frac{i}{2}[\gamma^\mu,\gamma^\nu]. 
\end{align}
Each form factor is further decomposed into 
the tree, 1PI and counterterm contributions as
\begin{align}
\Gamma_{Z'ff}^i = \Gamma_{Z'ff}^{i,{\rm tree}} + \Gamma_{Z'ff}^{i,{\rm 1PI}} + \delta\Gamma_{Z'ff}^i, ~~(i=V,~A,~M), 
\end{align}
where the tree-contribution is given by $\Gamma_{Z'ff}^{V,{\rm tree}} =v_f'$, $\Gamma_{Z'ff}^{A,{\rm tree}} =a_f'$ and $\Gamma_{Z'ff}^{M,{\rm tree}} = 0$ with $v_f'$ and $a_f'$ given in Eq.~(\ref{eq:coup-Zprime-ff}). 
The counterterm is given by 
\begin{align}
\delta\Gamma_{Z'ff}^V &= v_f'\left(\frac{\delta g_X}{g_X^{}} + \frac{\delta Z_{Z'}}{2} + \delta Z_V^f\right) + a_f' \delta Z_A^f, \\
\delta\Gamma_{Z'ff}^A &= a_f'\left(\frac{\delta g_X}{g_X^{}} + \frac{\delta Z_{Z'}}{2} + \delta Z_V^f\right) + v_f' \delta Z_A^f, 
\end{align}
where $\delta Z_V^f$ and $\delta Z_A^f$ are 
the wavefunction renormalization factors for a fermion $f$.
These are expressed as 
\begin{align}
\delta Z_V^f = \frac{\delta Z_L^f + \delta Z_R^f}{2},\quad 
\delta Z_A^f = \frac{\delta Z_L^f - \delta Z_R^f}{2},
\end{align}
with $\delta Z_{L}^f$ ($\delta Z_R^f$) being the wavefunction renormalization factors for a left-handed (right-handed) fermion.  
Due of the Ward-Takahashi identity, the 1PI contributions to the vector and axial-vector parts take the following form in the limit of $q^2 \to 0$: 
\begin{align}
\lim_{q^2 \to 0}\Gamma_{Z'ff}^{V,\text{1PI}} & = -v_f'\delta Z_V^f - a_f'\delta Z_A^f, \\ 
\lim_{q^2 \to 0}\Gamma_{Z'ff}^{A,\text{1PI}}  & = -a_f'\delta Z_V^f - v_f'\delta Z_A^f. 
\end{align}
Therefore, the renormalized form factors can be rewritten as 
\begin{align}
\hat{\Gamma}_{Z'ff}^V = v_f' + \Delta\Gamma_{Z'ff}^{V}(q^2) +   v_f'\left(\frac{\delta g_X}{g_X^{}} + \frac{\delta Z_{Z'}}{2} \right) , \\
\hat{\Gamma}_{Z'ff}^A = a_f' + \Delta\Gamma_{Z'ff}^{A}(q^2) +  a_f'\left(\frac{\delta g_X}{g_X^{}} + \frac{\delta Z_{Z'}}{2}\right), 
\end{align}
where $\Delta\Gamma_{Z'ff}^{X}(q^2) = \Gamma_{Z'ff}^{X,\text{1PI}}(q^2) - \Gamma_{Z'ff}^{X,\text{1PI}}(0)$ with $X = V,A$.

Now, we impose the renormalization condition as 
\begin{align}
\hat{\Gamma}_{Z'ff}^V(q^2)\Big|_{q^2 = 0} = v_f' , 
\end{align}
by which we obtain 
\begin{align}
\frac{\delta g_X^{}}{g_X^{}}  = -\frac{\delta Z_{Z'}}{2}. 
\label{eq:delvp2}
\end{align}
Combing Eqs.~(\ref{eq:delvp}) and (\ref{eq:delvp2}), we can determine the counterterm of the singlet VEV $\delta v_\Phi/v_\Phi$.

\end{appendix}
\bibliography{references}
\end{document}